\documentclass[11pt,a4paper]{article}
\usepackage{caltn,psfig}
\usepackage{graphicx}

\begin{document}

\docnum{IACHEC Report 2017}
\title{Summary of the 12th IACHEC Meeting}
\author{K.Forster$^a$, C.E.Grant$^b$, M.Guainazzi$^c$, V.Kashyap$^d$, H.L.Marshall$^b$, E.D.Miller$^b$, L.Natalucci$^e$, \\ J.Nevalainen$^f$, P.P.Plucinsky$^d$, Y.Terada$^g$ \\ \\
  ($^a$Cahill Center for Astronomy and Astrophysics, California Institute of Technology, USA \\
  $^b$Kavli Institute for Astrophysics and Space Research, Massachusetts Institute of Technology, USA \\
  $^c$ESA-ESTEC, The Netherlands \\
  $^d$Harvard-Smithsonian Center for Astrophysics, USA \\
  $^e$IAPS-INAF, Italy \\
  $^f$University of Tartu, Estonia \\
  $^g$University of Saitama, Japan
)
}
\date{\today}                   

\newcommand{\name}{1E~0102.2-7219}
\newcommand{\xmmn}{XMM-Newton}
\newcommand{\chan}{{\em Chandra}}
\newcommand{\suzaku}{{\em Suzaku}}
\newcommand{\swift}{Swift}
\newcommand{\nicer}{NICER}
\newcommand{\astrosat}{{\em Astrosat}}
\DeclareRobustCommand{\ion}[2]{\textup{#1\,\textsc{\lowercase{#2}}}}

\maketitle

\begin{center}
{\small
{\bf Abstract} \\
We summarize the outcome of the 12th meeting of
the International Astronomical Consortium for High Energy Calibration
(IACHEC), held at the UCLA conference center in Lake Arrowhead (California)
in March 2017.
56 scientists
directly involved in the calibration of operational and 
future high-energy missions gathered during 3.5~days to discuss the status of
the X-ray payload inter-calibration, as well as possible ways to improve it.
The ``Thermal Supernovas Remnant'' (SNR) Working Group presented a
recently published paper on 1E~0102.2-7219
as a calibration standard in the 0.5-1.0 keV band.
A new method to measure the high-energy spectrum of
the Crab Nebula and pulsar with NuSTAR without using its optics may yield a new
absolute flux standard in the 3--7~keV band.
A new ACIS contamination model - released with CALDB version 4.7.3 - leads to a significant
improvement in modeling the spectral, spatial, and temporal properties
of the contaminant.
The first calibration results of
the scientific payload on board {\it Hitomi} confirm the excellent performance
of the instruments before the spacecraft
operation problems leading to its loss.
Finally, the meeting discussed extensively a novel statistic approach to formally
identify in which direction the effective areas of different instruments
would need to be changed to bring them into concordance. This method
could inform future further
calibration efforts.
}
\end{center}

\section{Introduction}

The International Astronomical Consortium for High Energy Calibration
(IACHEC)\footnote{{\tt http://web.mit.edu/iachec/}}
is a group dedicated to supporting the cross-calibration
of the scientific payload
of high energy astrophysics missions with the ultimate goal
of maximizing their scientific return. Its members are drawn from
instrument teams, international and national space agencies, and other
scientists with an interest in calibration. Representatives
of over a dozen current and future missions regularly contribute to the
IACHEC activities. Support for the IACHEC in the form of travel costs
for the participating members is generously provided by the relevant
funding agencies.

IACHEC members cooperate within Working Groups (WGs) to define calibration
standards and procedures. The objective of these groups is primarily a
practical one: a set of data and results (eventually published in
refereed journals) produced from a coordinated and standardized
analysis of reference sources (``high-energy standard candles'').
Past, present and future high-energy missions can use these results as a
calibration reference.

The 12th IACHEC meeting was successfully hosted by the NuSTAR Team
based at the California Institute of Technology in Pasadena, CA, and was
held at the UCLA conference center in Lake Arrowhead, CA. The meeting
was attended by 56 scientists from the US, Europe, and Asia representing
high energy missions based in US, UK, Italy, Germany, France,
Netherlands, Switzerland, Spain, China, India, and Japan. The meeting
also featured presentations from IACHEC members unable to attend in
person who were able to view and contribute to the meeting via video
teleconferencing. Advances in the understanding of the calibration of
more than a dozen missions was discussed, covering multiple stages of
operation and development:

\begin{itemize}
  
\item[-] Recently completed missions - {\em Suzaku} and {\it Hitomi}
\item[-] Currently operating missions - {\em Chandra}, XMM-Newton, {\it Swift}, INTEGRAL,
{\em Astrosat}, POLAR, and NuSTAR
\item[-] Pre-launch status - NICER and HXMT (both launched in the summer of 2017)
and eROSITA
\item[-] Missions under development: IXPE, XARM, and {\it Athena}
\item[-] Future mission concepts: Cal X-1
\item[-] Ground calibration facilities: acceleration facilities at the IAAT of the University of T\"ubingen, GSFC/CREEST, MPE/Panter, NASA/XRCF among others
\item[-] Atomic databases: AtomDB

\end{itemize}

This report summarizes the main results of the 12th meeting and
comprises the reports from each of the IACHEC WGs. The presentations held
at the meeting are available at: \\
{\tt http://web.mit.edu/iachec/meetings/2017/index.html}.

The IACHEC gratefully acknowledges sponsorship for the meeting that was
provided by the NuSTAR mission, the Chandra X-ray Center at the
Smithsonian Astrophysical Observatory, and the NASA Goddard Space Flight
Center. Sponsorship was also provided by the Committee on Space Research
(COSPAR) to cover the cost of meeting registration for two participants
from the {\em Astrosat} mission who would not otherwise have been able to
attend.

\section{Working Group reports}

\subsection{Calibration Uncertainties}

The Calibration Uncertainties WG changed its goals and expanded to become a forum
for the discussion of statistical, methodological, and algorithmic issues that affect the calibration of astronomical instruments and how they are used in data analysis and interpretation of analysis results.

\subsubsection{XSPEC}

During the working group meeting, Keith Arnaud presented planned updates to XSPEC that might be especially useful for calibration analysis:
\begin{enumerate}
\item {\bf {\tt mdefine}} -- A flexible way to allow algebraic modifications to model library
\item {\bf Multiple responses} - A method to allow a set of responses to be used with a single dataset, e.g., break a grating RMF into high-resolution (mainly diagonal with few non-zero elements) and smaller, low-resolution (triangular to describe the tail down to low energies) RMFs
\item {\bf Modified RMF} --  Possibly include a new format for the
Redistribution Matrix File (RMF) developed by Kaastra \& Bleeker (2016) which prescribes an average energy as well as a slope in each bin.  The consensus of the WG was that it is not yet clear whether this type of RMF definition is optimized for a particular type of spectrum, whether the effect of Gaussian assumptions would translate well to the Poisson regime in which spectra are used, and whether the performance improvements promised by the method would be relevant in five years when XARM would fly.
\end{enumerate}

\subsubsection{Concordance}

A group of astronomers and statisticians are developing a method to formally identify how much and in which direction the effective areas of different instruments need to change to bring them into concordance.

The method is based on shrinking the disparate measurements from many instruments observing many common sources such that the information about the source fluxes are pooled together to gain insight into how discrepant each instrument is relative to an asymptotically perfect estimate.

Herman Marshall gave a plenary talk on Tuesday, Mar 28 introducing the concept and what it would take to achieve this goal. The key was that each instrument team would specify what they believe to be the magnitude of the systematic error in their calibration, and these numbers would be used, in conjunction with appropriate measurements in different pass-bands, to infer the corrections. These inferences are expected to guide the instrument teams in zeroing in on where their calibration should be changed.

A great many of these systematic error estimates, $\tau$, were collected during the Roundtable discussion on Wednesday, Mar 29. Many more instrument teams have promised to supply well-thought out numbers shortly. The next stage of the concordance effort will be to start gathering suitable datasets. The immediate goal of the Concordance Project members is to publish the method in a Statistics journal, followed by exemplar applications to be published in an Astronomy journal.

Additionally, they plan to account for various complicating effects like gracefully handling measurement outliers using the log-t distribution in place of a log-Normal, as discussed in detail by Xufei Wang during the WG meeting.  There are plans to include a measure of cross-band covariance from
Jeremy Drake's Monte Carlo effective area generating process (MCCal)
such that information can be usefully borrowed from measurements made in multiple pass-bands.

\subsection{CCD}

The CCD and Backgrounds WG met in two well-attended sessions.  As always, the CCD Working Group provides a forum for cross-mission
discussion and comparison of CCD-specific modeling and calibration issues, while the Backgrounds WG provides the same for measuring and
modeling instrument backgrounds in the spatial, spectral and temporal dimensions.  Attendees represented all the relevant CCD-using X-ray missions,
including XMM-Newton, {\it Chandra}, {\it Hitomi}, and {\it Swift}.

The first session began with a talk given remotely by Silvano Molendi, ``An in-depth analysis of the EPIC-MOS instrumental background.''  This work is
part of a large effort to better understand the XMM-Newton particle background, the particle environment in High-Earth Orbit and L2, and the expected
background for {\it Athena}. A number of related papers have appeared in SPIE proceedings and in an upcoming issue of Experimental Astronomy, expected
summer 2017.  They have undertaken a systematic analysis of the XMM-Newton EPIC background data over the full mission. EPIC-MOS data from inside and outside the field-of-view of the mirrors is used to separate the focused and unfocused components of the background.  The spectral and temporal behavior of the background is then compared to the magneto-spheric and orbital position of XMM-Newton

We then heard from Konrad Dennerl on ``An empirical method for improving the XMM-Newton\\/EPIC-pn RMF and ARFs.''  He described a project to parameterize
and optimize redistribution matrix (RMF) components and showed a proof of concept with a few soft X-ray ($\le$2~keV) sources, and significant improvements.  He plans to extend
the work to higher energies, other readout modes, time periods, and detector positions.

Terry Gaetz and Nick Durham discussed three Chandra ACIS calibration topics: ``{\it Chandra} ACIS Background,'' ``{\it Chandra} ACIS I3 Response Width,''
and
``ACIS Gain Studies: Temporal, Spatial, and Temperature Dependencies.'' A new set of blank sky background data covering the period 2012 through 2015 is
nearly ready for release.  Looking further into the future, improvements for the ACIS-I3 response products with better time and temperature dependence
are underway.  Finally, the time-dependent gain calibration of ACIS CCDs is being reviewed with an eye toward improvements in the process and the
results.

Koji Mori gave a comprehensive talk on the the ``{\it Hitomi} Soft X-ray Imager (SXI) overview and on-board calibrations.''
The SXI provided a wide field of
view (38'$\times$38') and high Quantum Efficiency from 0.4--12~keV and was the first P-channel X-ray CCD in space. He reported on some on-orbit
surprises and on the non X-ray background.

Finally, Andy Beardmore presented ``XRT Windowed Timing (WT) Mode Trailing Charge.'' The {\it Swift} XRT WT mode low-energy background is increasing
with time.
This increase has been identified with trailing charge released from deep charge traps.  The XRT team is developing an algorithm to identify and
remove trailed events.

\subsection{Contamination}

The Contamination WG met for one session of updates from recently operating missions.  As at previous IACHEC workshops, the group discussed effects of molecular contamination on soft X-ray
instruments (e.g., Marshall et al. 2004, Koyama et al. 2007, O'Dell et al. 2013). Since its inception, the WG has covered three broad topics: (1) comparison of contamination among
instruments and missions; (2) mitigation for current instruments; and (3) mitigation for future instruments.

Much of the session was devoted to updates on {\it Chandra}
ACIS contamination monitoring and characterization. Akos Bogdan presented measurements of the time and spatial evolution of the ACIS contaminant
using raster observations of Abell~1795, showing that the optical depth is increasing rapidly at recent times (Plucinsky et al. 2016). Herman Marshall presented a physical chemical model of the
contaminant, using the high-spectral-resolution LETGS in several `big dither' observations of bright blazars.  The LETGS resolves structure of the C-K resonance feature, allowing separate modeling of
aliphatic and aromatic components and providing important data to inform mitigation plans. The newly released ACIS contamination model has already been used to great success by members of the
WG (CALDB~4.7.3).
Eric Miller and Koji Mori presented two {\it Hitomi} SXI observations of the isolated neutron star RXJ1856 taken a week apart, obtaining $N_{\rm C} \leq 9.4\times10^{17}$ carbon atoms cm$^{-2}$ as an upper limit of contamination on the instrument. Steve Sembay presented an update on the XMM-Newton EPIC-MOS contamination, which is relatively small, different between MOS1 and MOS2, and shows a similar trend to previous years.
He noted again that EPIC-pn, which has a cold trap, shows no sign of contamination.

The plan for compiling a legacy white paper including ``lessons learned'' and advice for future missions was debated since there was a concern among some of the attendees that these topics are too broad
for a single working group.  Going forward, the Contamination WG will focus on comparison, monitoring, and mitigation among current and past instruments (topics [1] and [2] above).
Mitigation for future instruments is tentatively planned for incorporation into a new Optics WG. Since much of the recent contamination analysis is
only published in not refereed proceedings,
technical memos, and user documentation, calibration scientists from several missions indicated their plans to submit manuscripts to refereed journals over the course of the next year.

\subsection{Coordinated observations}

In the Coordinated Observations WG, we heard the status of several on-going projects involving campaigns where several X-ray observatories were
coordinating observations of a particular target.  In the past year, a group of IACHEC calibration scientists led by
Kristin Madsen published results on campaigns carried out in 2012 and 2013 on
PKS~2155-304 and 3C 273, respectively, involving NuSTAR, {\it Chandra}, {\it Swift}, XMM-Newton, and {\it Suzaku} (Madsen et al. 2017a). Kristin Madsen will
also check to see if her scripts can be easily adapted to apply to observations of 3C~273 which were obtained in 2015 and 2016.  She also reported on a new,
very interesting method to measure the high energy flux of the Crab Nebula and pulsar by viewing it without using the NuSTAR optics
(Madsen et al. 2017b).  This method could yield a new absolute flux standard in the 3-7~keV band that would imply adjustments are needed in other
instruments' effective areas in this band.  In a related presentation, Lorenzo Natalucci presented the 2012 and 2016
observations of 3C~273, finding that INTEGRAL
and NuSTAR agree well in their overlapping energy range.  The XMM-Newton/{\it Chandra} blazar project is still on hold due to problems with
XMM-Newton data analysis correcting for PSF clipping to avoid pile-up.

The WG discussed what joint observations may be made with NICER and HXMT, which should be available for
cross-calibration observations in the September/October time frame.  There are several campaigns that should work well for NICER, such as the campaign on
3C~273 that is already scheduled for many missions in June, Mkn~421 in June that will be observed by {\it Chandra} for two days, and a possible joint
observation of the Crab Nebula with NuSTAR in August.  Coordinating an observation of SS~433, with high energy lines that are good for calibration,
may be possible with {\it Chandra} using a Director Discretionary Time observation.  For HXMT, observations of bright Galactic targets are best and
may be possible jointly with NuSTAR and NICER.  The {\it Astrosat} team will work on joining the coordinated observations campaign on 3C~273 in 2018.

Kristin Madsen raised an issue about accounting for dust scattered halos around distant Galactic sources that can give disparate flux measurements if a
timing mode is used. A joint observation with XMM-Newton has been planned to address this issue.

\subsection{Galaxy Clusters}

We continued planning the Multi Mission Study project whereby we aim at
comparing X-ray spectroscopic results of a sample of clusters obtained
with five on-going and past X-ray missions and nine instruments. In particular, we
discussed in detail the criteria for suitable clusters with feasible
exposure times to achieve our desired statistical precision of $\sim$1\% in $\sim$10
spectral bins in the 0.5--7~keV band.

Given the large variability of the effective areas of different
instruments, it is very difficult to build a common cluster sample with
statistically meaningful number of members, with observations reaching our
pre-defined precision level. After discussion we decided upon our criteria
for the data quality and cluster properties.

We agreed to the time table that by the end of April we will have
completed the archival study of the suitable clusters and observations. By
end of June we will have the data processed. By the end of the year Jukka Nevalainen
will do the stack residual analysis, and prepare a draft of the paper.

\subsection{Heritage}

The Heritage Working Group aims at: a) providing a platform for the discussion of experiences coming from operational missions,
b) facilitating the usage of good practices for the management of pre- and post-flight calibration data and procedures, and the maintenance and
propagation of systematic uncertainties (the latter task in strict collaboration with the ``Calibration Uncertainties'' IACHEC WG),
c) documenting the best practices in analyzing high-energy astronomical data as a reference for the whole scientific community,
d) ensuring the usage of homogeneous data analysis procedures across the IACHEC calibration and cross-calibration activities,
e) consolidating and disseminating the experience of operational missions on the optimal calibration sources for each specific calibration goal.

The main current activities of this WG are:

\begin{itemize}

\item Making available calibration-related papers and documents of past and operational missions
  through the WG Wiki{\footnote {\tt https://wikis.mit.edu/confluence/display/iachec/IACHEC+Heritage+Working+Group}}

\item Implementing a database of observational data used by the IACHEC for calibration and cross-calibration studies. A first version of this
  database, including data employed in the IACHEC paper published so far, is expected to be available by the next yearly IACHEC meeting in 2018

\end{itemize}

\subsection{Non-Thermal SNRs}

The session of the Non-Thermal SNR WG was devoted to discussing the status of the Crab cross-calibration project and the planning of a future paper on G21.5-0.9, taking into account the NuSTAR results and possibly including the new {\it Hitomi} data. Recent results from Fermi/GBM reported a decline in the overall Crab flux since MJD 57000, consistent with BAT, down to the level of the minimum previously observed in 2011.

Monitoring programs of the Crab from other satellites are ongoing: INTEGRAL performs bi-monthly observations during its visibility periods (covering $\sim4$~months/year). A proposal for a monthly campaign was submitted for NuSTAR with the aim to measure $\sim2$\% flux variations and get accurate snap-shot spectra.

The Crab cross-calibration project in the hard X-rays takes advantage of a big dataset with 14 nearly-simultaneous epochs from 2005-2014, plus an observation of XMM-Newton in burst mode. A new dataset from BAT has been now added (8 epochs). Results from the different epochs are nearly consistent in terms of flux normalization differences. The current NuSTAR normalization is $\sim10-12$\% lower than INTEGRAL and $\sim20-25$\% lower than PCA and HXD/PIN. However, it looks like the absolute flux measured with NuSTAR with the Crab in stray-light position essentially agrees with the INTEGRAL measurement. At the meeting it was decided to add a nearly simultaneous epoch in 2016 (25 March to April 1) including NuSTAR, INTEGRAL and {\em Hitomi} data (it was noticed that {\em Astrosat} has also observed the Crab in the same period). NuSTAR is now observing the Crab on a monthly basis in the framework of its
calibration program.

A proposal for a new paper on G21.5-0.9 (with a focus on $>$5 keV band) was discussed. The motivation for this work are: the availability of new data (for example: NuSTAR, {\em Hitomi}, INTEGRAL), improvements in the calibration (e.g., hard band response for {\it Suzaku}/XIS)
and a new model derived from NuSTAR observations (energy break near 10 keV). {\em Hitomi}/SXS spectra of G21.5-0.9 were presented.

\subsection{Thermal SNRs}

\begin{figure*}
 \begin{center}
 \includegraphics[width=11.0cm,angle=90]{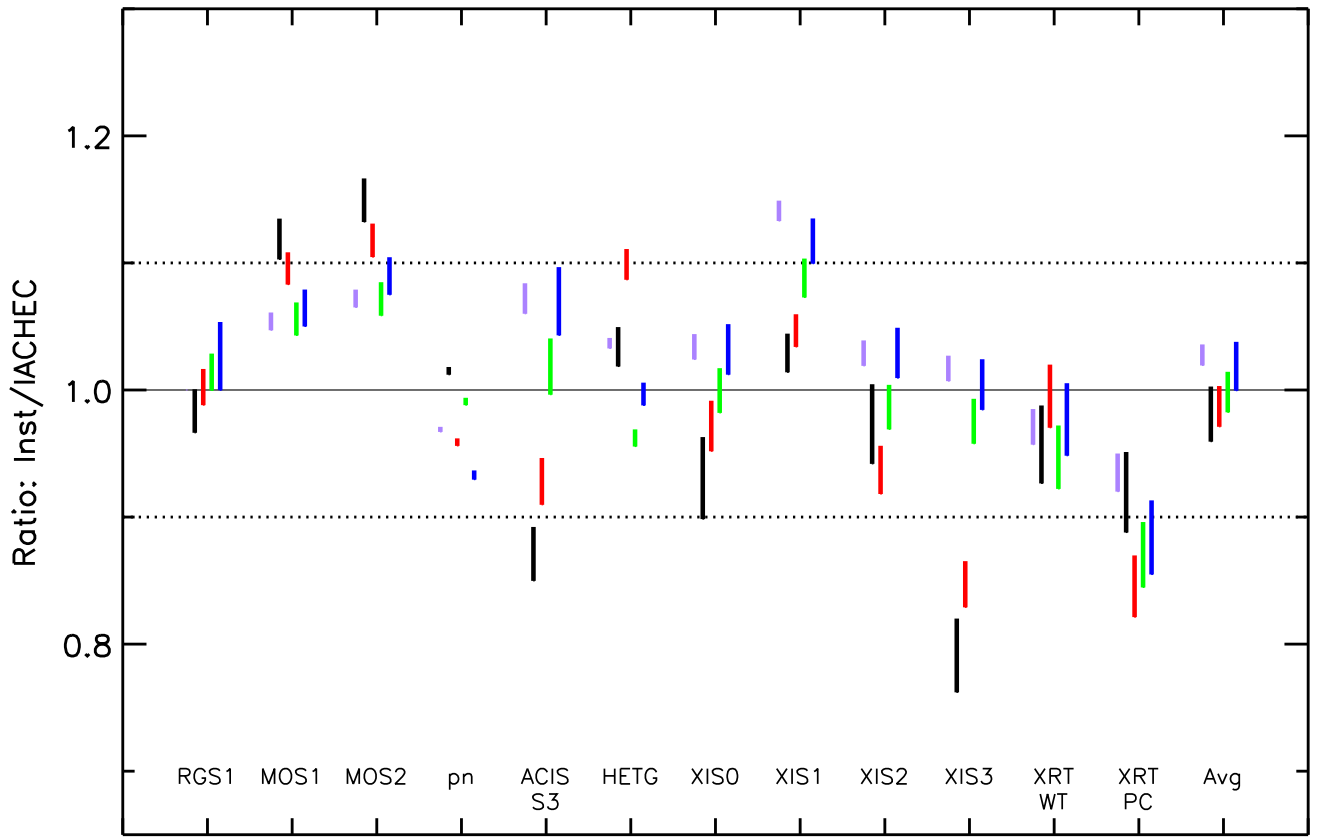}
 \end{center}
 \caption{\label{fig:comp_norms} {Comparison of the scaled
 normalizations for each instrument to the IACHEC model values and the average.
 There are four or five  points for each
instrument which are from left to right, global normalization (purple),
\ion{O}{vii}~He$\alpha$~{\em r} (black), \ion{O}{VIII}~Ly$\alpha$ (red), \ion{Ne}{ix}~He$\alpha$~{\em r} (green), and \ion{Ne}{X}~Ly$\alpha$ (blue). The
length of the line
indicates the $1.0\sigma$ confidence limit for the scaled normalization.
 } }
\end{figure*}

The Thermal SNR Working Group completed work on the IACHEC paper on 
cross-calibration using 1E~0102.2-7219 (hereafter E0102) and the paper has 
appeared in Astronomy and Astrophysics (Plucinsky et al. 2017).  We used the IACHEC standard model for E0102 to fit the data from the
current generation of X-ray CCD instruments, specifically: \chan\/ ACIS-S3, 
\xmmn\/ (EPIC-MOS and EPIC-pn), \suzaku\/ XIS, and \swift\/ XRT.
We performed our effective area comparison with representative, early
mission data when the radiation damage and contamination layers were
at a minimum, except for the \xmmn\/ EPIC-pn instrument which is
stable in time.   We found that the measured fluxes of the 
\ion{O}{vii}~He$\alpha$~{\em r} line, the \ion{O}{viii}~Ly$\alpha$ line, the 
\ion{Ne}{ix}~He$\alpha$~{\em r} line, and the \ion{Ne}{X}~Ly$\alpha$ line 
generally agree to within $\pm10\%$ for all instruments, with 38 of our 
48 fitted normalizations within $\pm10\%$ of the IACHEC model value. 
Figure~\ref{fig:comp_norms} displays a comparison of the fitted line 
normalizations with respect to the standard IACHEC model.
We then fit all available observations of E0102 for the CCD instruments close 
to the on-axis position to characterize the time dependence in the 
0.5--1.0~keV band. We present the measured line normalizations as a function 
of time for each CCD instrument so that the users may estimate the uncertainty
 in their measured line fluxes for the  epoch of their observations.

The Thermal SNR Working Group met once during the 2017 IACHEC meeting with 
the following attendees contributing in person: Eric Miller (\suzaku~XIS), 
Steve Sembay (\xmmn~MOS), Andy Beardmore (\swift~XRT), Martin Stuhlinger
(\xmmn~calibration), Adam Foster (Models), Sunil Chandra (\astrosat), and 
Paul Plucinsky (\chan~ACIS), and the following contributing remotely:  Frank 
Haberl (\xmmn~pn), and Andy Pollock (\xmmn~RGS). There were several other
conference attendees who sat in on the session. The group continues to use
the standard IACHEC model for E0102 and newly acquired observations of E0102 
to test and verify the calibrations for their respective instruments.  Frank 
Haberl showed the pn results with the latest version of the \xmmn\/ software, 
SAS16. He also commented that Konrad Dennerl is working on modifications to
the pn response matrix with the hope of improving the fits to the O lines (cf. the ``CCD WG'' Section in this report).  
Paul Plucinsky showed the on-axis and off-axis results with the new ACIS contamination 
model (N0010) that demonstrate that the spatial variation in the thickness of 
the contamination layer on ACIS is characterized better in the new model. 
Sunil Chandra showed the {\astrosat} fits to E0102, which are a significant 
improvement compared to the fits shown at the last IACHEC meeting due to a 
modified gain calibration.  One issue for {\astrosat} is the size of the 
extraction region and contamination from a nearby Be XRB.  Sunil Chandra showed 
results with extraction regions of 8, 13, and 18 arc-minutes.

 The main focus of the WG at this meeting was the development of a standard 
IACHEC model for N132D, the brightest SNR in the Large Magellanic Cloud which 
has a significantly different spectrum than E0102.  N132D has multiple plasma
components contributing to the overall X-ray spectrum with higher temperatures than in
E0102, resulting in emission lines of Si, S, Ar, Ca, and Fe-K. The WG intends
to construct a model which could be used to determine the normalizations of
the Si, S, and Ar lines in the 1.5 to 3.5 keV band and perhaps the Fe-K line
at 6.7 keV. The current version of the IACHEC model for N132D is referred to 
as v2.10.  This version of the model uses the latest release (v3.0.8) of the
APEC data, uses the {\tt nlapec} model in XSPEC, and removes the power-law
component which was intended to be a crude background model.  Each instrument
scientist is now responsible for modeling the background for their
respective instrument.  Paul Plucinsky showed fit results with v2.10 of the model to 
representative {\xmmn}\/ MOS and pn spectra.  There were issues around the 
\ion{Si}{xiii} lines that appear to be related to the gain calibration for 
these data sets.  Eric Miller showed fits to the XIS spectra with v2.10.  There were 
also issues with the fit around the \ion{Si}{xiii} lines but most of this 
appears to be related to known issues with the XIS energy scale at the Si 
edge. Andy Pollock showed a comparison of the lines in v2.9 of the model 
compared to a representative RGS data set.  He commented that the 
\ion{O}{viii}~Ly$\alpha$ line showed signs of pile-up in the RGS data, and
suggested that he stack all of the available RGS data on N132D to improve the 
characterization of the weak lines in the spectrum.  The WG was hesitant 
about this idea since the CCD instruments are relatively insensitive to the 
weak lines and a thorough job of characterizing the weak lines could be a 
time-consuming task.  The WG suggested that Andy Pollock stack a small set of RGS 
observations early in the mission and finalize the line characterization with 
those data.  Steve Sembay and Andy Beardmore wondered why we are fitting with the 
continuum frozen when the continuum is significant, perhaps even stronger, 
compared to the lines we might hope to fit.  The WG will experiment with the 
normalization and shape of the continuum in the broad band and the narrow 
band (1.5-2.5 keV) to understand the effects on the fits.  The goal of the WG 
is to finalize the standard IACHEC model for N132D well in advance of the 
next IACHEC meeting so that fit results with that model can be compared during
that meeting.

\subsection{Timing}

On the timing WG, we shared the current status of the time assignment on {\em Suzaku} and {\it Hitomi} satellites, reported by S. Koyama, Y.Terada, and K.Oshimizu. On {\em Suzaku} time assignment, they found an anomaly on the time assignment system on the ground station with the timing observations of Crab pulsar and tried to recover the original timing accuracy at about 300~$\mu$sec (Terada et al. 2007). On {\it Hitomi} satellite, they reviewed an overall structure of the timing system of the spacecraft and the algorithms of the off-line time-assignment tools, and reported how they control the error budget for timing accuracy on each component. They also reported the results of actual measurements on timing on ground and in-orbit. The timing accuracy of the
Modulated X-ray Source on ground timing measurements was discussed between {\it Hitomi} and NICER missions etc.

After several discussions, we decided to continue this WG for near-future X-ray missions. Historically, the timing WG began from the 2nd IACHEC at UCLA on 2007 but closed on the 4th meeting in Japan because the goals are satisfied on missions at that moment, and resumed from the 9th IACHEC on 2014 for
{\it Hitomi} (at that time still ASTRO-H), NuSTAR, and NICER etc. Three goals are defined again for the current timing WG: 1) to share information among various missions on the methods and procedures of the timing calibration both in orbit and on ground and the lessons learned from the previous missions, 2) to trigger the future in-orbit timing calibration, and 3) to proceed basic timing studies on such as effects of dead-time, grade selection on the products.

\section*{References\footnote{see {\tt http://web.mit.edu/iachec/papers/index.html} for a complete list of IACHEC papers}}

\noindent
Kaastra J. \& Bleeker J.A.M., 2016, A\&A, 587, 151 \\
\noindent
Koyama Y., et al., 2007, MNRAS, 382, 1719\\
\noindent
Madsen K., et al., 2017a, AJ, 153, 2 \\
\noindent
Madsen K., et al., 2017b, ApJ, 841, 56 \\
\noindent
Marshall H., et al., 2004, SPIE, 5165, 497 \\
\noindent
O'Dell S., et al., 2013, SPIE, 8559, 0 \\
\noindent
Plucinsky P., et al., 2016, SPIE, 9905, 44 \\
\noindent
Plucinsky P., et al., 2017, A\&A, 597, 35 \\
\noindent
Terada Y., et al., 2008, PASJ, 60, 25

\end{document}